\documentclass{sig-alternate}

\usepackage{color}

\newcommand{\ignore}[1]{}

\begin{document}

\title{Afterburner: The Case for In-Browser Analytics}

\numberofauthors{2}
\author{
Kareem El Gebaly and Jimmy Lin\\[1ex]
\affaddr{David R. Cheriton School of Computer Science}\\
\affaddr{University of Waterloo, Ontario, Canada}\\[1ex]
\affaddr{\{kareem.elgebaly, jimmylin\}@uwaterloo.ca}
}

\maketitle
\begin{abstract}
This paper explores the novel and unconventional idea of implementing
an analytical RDBMS in pure JavaScript so that it runs completely
inside a browser with no external dependencies. Our prototype, called
Afterburner, generates compiled query plans that exploit typed arrays
and asm.js, two relatively recent advances in JavaScript. On a few
simple queries, we show that Afterburner achieves comparable
performance to MonetDB running natively on the same machine. This is
an interesting finding in that it shows how far JavaScript has come as
an efficient execution platform. Beyond a mere technical curiosity, we
discuss how our techniques could support ubiquitous in-browser
interactive analytics (potentially integrating with browser-based
notebooks) and also present interesting opportunities for
split-execution strategies where query operators are distributed
between the browser and backend servers.
\end{abstract}

\section{Introduction}

Browser-based notebooks (i.e.,~Jupyter) have gained tre\-men\-dous
popularity with data scientists in recent years for a variety of
reasons:\ The tight integration of code and execution output elevates
the analytical process and its products to first class citizens, since
the notebook itself can be serialized, reloaded, and shared. The
ability to manipulate, rearrange, and insert snippets of code (in
``cells'') lines up well with the iterative nature of data science and
a wide range of analytics tasks. A recent development is the seamless
integration of browser-based notebooks with scalable data analytics
platforms, such that code written in a notebook cell can be executed
on a potentially large cluster and the results can be further
manipulated in the notebook. This integration, exemplified by the
commercial Databricks platform that provides a notebook frontend to
Spark clusters~\cite{Zaharia_etal_NSDI2012}, allows data scientists to
analyze large amounts of data in a convenient and flexible manner.

The advent of notebooks means that the browser has, in essence, become
the ``shell''. However, in all implementations that we are aware of,
the browser is simply a dumb rendering endpoint:\ all query execution
is handled by backend servers. In the terminology of Franklin et
al.~\cite{Franklin_etal_SIGMOD1996}, this is referred to as {\it query
  shipping}.  However, modern browsers are capable of so much
more:\ they embed powerful JavaScript engines capable of running
real-time collaborative tools, online multi-player games, rendering
impressive 3D scenes, supporting complex, interactive visualizations,
and even running first-person shooters. These applications take
advantage of HTML5 standards such as WebGL, WebSocket, and
Indexed\-DB, and therefore do not require additional plug-ins
(compared to Flash or, yuck, Java applets).

We asked ourselves:\ Is it possible to exploit modern JavaScript
engines and build a high-performance data management system that runs
{\it completely in the browser}? And if so, what new opportunities
does such a platform create? We tackle these two questions in
turn:\ First, we present Afterburner, a prototype analytical RDBMS
implemented in JavaScript that executes completely inside a web
browser, is standards compliant, and has no external dependencies.
Experiments with a few simple queries on a modest dataset show that
Afterburner achieves comparable performance to the analytical database
MonetDB~\cite{Boncz_etal_CIDR2005} {\it running natively on the same
  machine}.  This finding is interesting in that it shows how far
JavaScript has come as an efficient execution platform, from its
much-maligned performance characteristics in the early days.  Second,
we discuss the ``so what?''\ question by highlighting the potential of
our techniques for supporting ubiquitous in-browser interactive
analytics and query execution strategies that make use of query
operators running both in the browser and on backend servers.

The primary contribution of our work is a feasibility demonstration of
in-browser analytics. To our knowledge, we are the first to propose
this somewhat unconventional idea of embedding an analytical RDBMS
inside the browser and exploring the implications thereof.  The idea
of splitting query execution across the client and the server, of
course, is not new, but we believe that JavaScript presents a fresh
take on the decades-old technique.

\section{Afterburner Implementation}

Afterburner is implemented as a JavaScript library, primarily designed
to run inside a standards-compliant web browser. However, with minimal
modifications it can run in other JavaScript environments such as
node.js.  In the current implementation, data is loaded into the
browser from a flat file on the client's file system, although in
principle data ingest could be accomplished via a REST API call, a
WebSocket, or a variety of other means. All data are immutable and
packed in a columnar layout in memory once loaded.  Afterburner
generates compiled query plans that exploit two JavaScript
features:\ typed arrays and asm.js, which we explain below.

\subsection{In-Memory Storage}


\begin{figure}[t]
  \includegraphics[width=.48\textwidth, keepaspectratio]{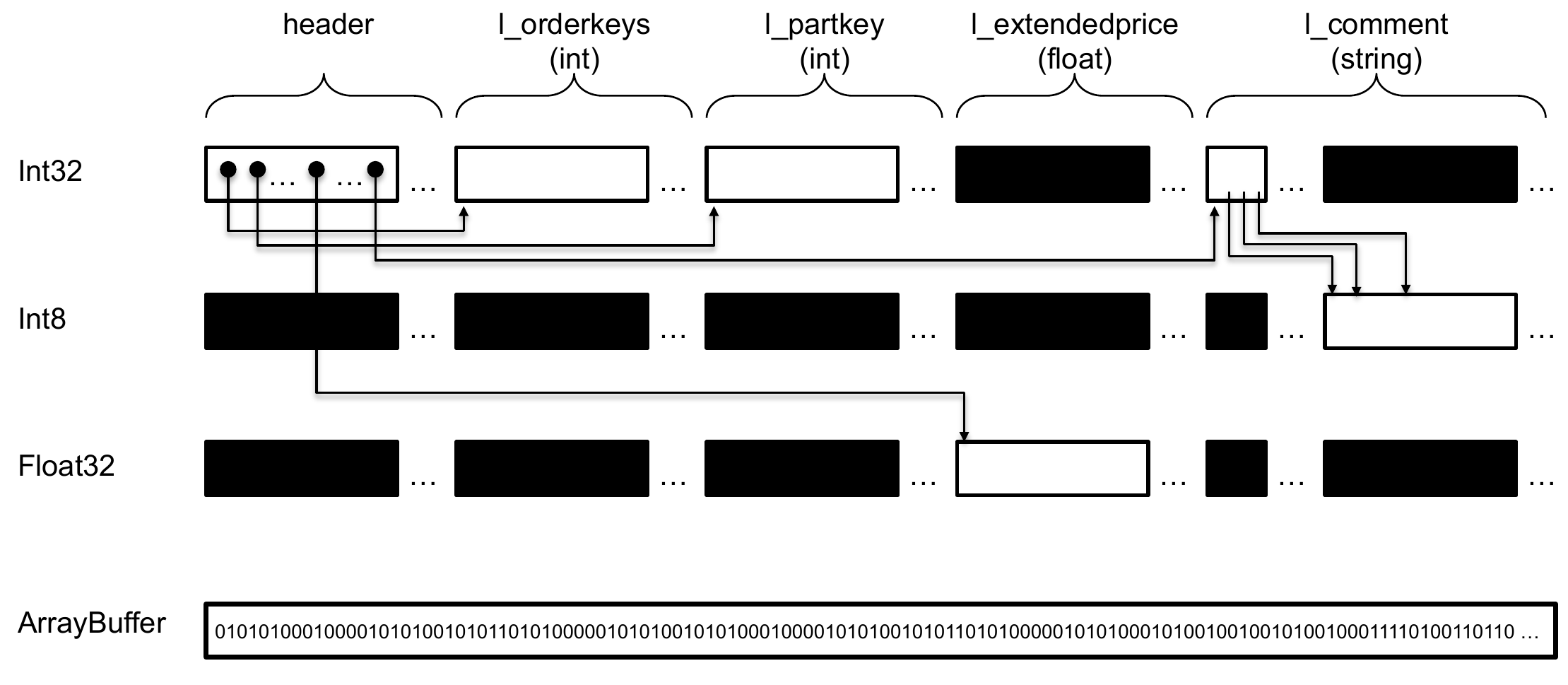}
  \caption{Illustration of the physical in-memory representation of
    the {\tt lineitem} table from TPC-H. Different ``views'' (top
    three rows) provide access into the underlying JavaScript
    ArrayBuffer. Blackened boxes represent invalid data for that
    particular view.}
  \label{fig:memory}
\end{figure}

Array objects in JavaScript can store elements of any type and are not
arrays in a traditional sense (compared to say, C) since consecutive
elements may not be contiguous; furthermore, the array itself can
dynamically grow and shrink. This flexibility limits the optimizations
that the JavaScript engine can perform both during compilation and at
runtime. In the evolution of JavaScript, it became clear that the
language needed more efficient methods to quickly manipulate binary
data: typed arrays are the answer.

Typed arrays in JavaScript are comprised of buffers, which simply
represent untyped binary data, and views, which impose a read context
on the buffer. As an example, the following creates a 64-byte buffer:

\begin{small}
\begin{verbatim}
  var buffer = new ArrayBuffer(64);
\end{verbatim}
\end{small}

\noindent Before we can manipulate the data, we need to create a view
from it. With the following:

\begin{small}
\begin{verbatim}
  var int32View = new Int32Array(buffer);
\end{verbatim}
\end{small}

\noindent we can now manipulate {\tt int32View} as an array of 32-bit
integers (e.g., iterate over it with a {\tt for} loop).

Typed arrays allow the developer to create multiple views over the
same buffer, which provides a mechanism to interact with, for example,
arbitrary C {\tt struct}s or complex WebGL data. Afterburner takes
advantage of exactly this feature to pack relational data into a
columnar layout.

In Afterburner, each column is laid out end-to-end in the underlying
buffer (which can be traversed with a view of the corresponding
type). The table itself is a group of pointers to the offsets of the
beginning of the data in each column.  Figure~\ref{fig:memory} shows
the physical memory layout storing the {\tt lineitem} table from the
TPC-H benchmark, which we use as a running example.  A {\tt lineitem}
pointer serves as the entry point into a group of 32-bit integer
pointers, which represent the offsets of the data in each column ({\tt
  l\_orderkeys}, {\tt l\_partkeys}, etc.).  Currently, our
implementation supports integers, floats, dates, and strings.  For the
first three types, values are stored as literals (essentially, as an
array). For a column of strings, we store null-terminated strings
prefixed with a header of pointers into the beginning of each string,
essentially a {\tt (char **)} in C.

Intermediate data for query execution in Afterburner are also stored
using typed arrays.

\subsection{Query Compilation}


In conjunction with typed arrays, Afterburner takes advantage of
asm.js, a strictly-typed subset of JavaScript that is designed to be
easily optimizable by an execution engine. Consider the following
fragment of JavaScript for counting the number of records that matches
a particular predicate on the column {\tt extendedprice}:

\begin{small}
\begin{verbatim}
  function count(val){                  
    var cnt = 0;
    for (var id; id < orderkey.length; id++)
      if (extendedprice[id] < val) cnt++;
    return cnt;
  }                                      
\end{verbatim}
\end{small}

\noindent For expository convenience, we refer to JavaScript without
asm.js optimizations as {\it vanilla} JavaScript. The equivalent
function in asm.js is as follows:

\begin{small}
\begin{verbatim}
  function count_asm(val, length){
    "use asm";
    val=+(val);
    length=length|0;
    length=length<<2;
    id=0;
    while ((id|0) < (length|0)){
      if (+(extendedprice[id>>2]) < +(val))
        cnt=(cnt+1)|0;
      id=(id+4)|0;
    }
    return cnt|0;
  }
\end{verbatim}
\end{small}

\noindent A key feature of asm.js is the use of type hints, such as
{\tt x|0} and {\tt +(x)}, which are applied to variables or
arithmetic expressions. The type hint ({\tt x|0}) specifies a 32-bit
integer and {\tt +(x)} specifies a 32-bit floating point value.  With
these hints, asm.js essentially introduces a static type system while
retaining backwards compatibility with vanilla JavaScript, since in
vanilla JavaScript these hints just become no-ops.  Note that accesses
to typed views must indicate the byte offset and the size of each
element.  The {\tt extendedprice} column is a 32-bit float view and
thus the byte offsets can be computed by multiplying the index variable
{\tt n} by 4 using the shift operator ({\tt >{>}}).

Any JavaScript function can request validation of a block of code as
valid asm.js via a special prologue directive, {\tt use asm}, which
happens when the source code is loaded.  Validated asm.js code
(typically referred to as an asm.js module) is amenable to
ahead-of-time (AOT) compilation, in contrast to just-in-time (JIT)
compilation in vanilla JavaScript. Executable code generated by AOT
compilers can be quite efficient, through the removal of runtime type
checks (since everything is statically typed), operation on unboxed
(i.e., primitive) types, and the removal of garbage collection.

An asm.js module can take three optional parameters, which
provide hooks for integration with external vanilla JavaScript
code:\ a standard library object, providing access to a limited subset
of the JavaScript standard libraries; a foreign function interface
(FFI), providing access to custom external JavaScript functions; and a
heap buffer, providing a single ArrayBuffer to act as the asm.js
heap.\footnote{\tt http://asmjs.org/spec/latest/} Thus, a typical
asm.js module declaration is as follows:

\begin{small}
\begin{verbatim}
  function MyAsmModule(stdlib, foreign, heap) {
    "use asm";

    // module body
  }
\end{verbatim}
\end{small}

At a high-level, Afterburner translates SQL into the string
representation of an asm.js module (i.e., the physical query plan,
through code templates described below), calls {\tt eval} on the code,
which triggers AOT compilation and links the module to the calling
JavaScript code, and finally executes the module (i.e., executes the
query plan).  The typed array storing all the tables (i.e., the entire
database) is passed into the module as a parameter, and the query
results are returned by the module.

The compiled query approach of Afterburner takes after systems like
HIQUE~\cite{hiq} and HyPer~\cite{Neumann11}, which have recently
popularized the idea of code generation for relational query
processing. One well-known drawback of this approach is that compiling
generated code using a tool like {\tt gcc} can overshadow its benefits
for short-running queries. As an alternative, Neuman~\cite{Neumann11}
proposed a hybrid compilation model, where the queries are compiled
into an intermediate representation such as LLVM, which is then linked
to pre-compiled native code to achieve native-like performance while
avoiding native compilation overhead. In the context of our work, we
have found compilation overhead to be negligible, primarily because
compilation speed is already something that JavaScript engines
optimize for, since all JavaScript code on the web is stored as text.

\subsection{Code Generation}

Instead of string-based SQL queries, Afterburner executes queries
written using an API that is heavily driven by method chaining, often
referred to as a {\it fluent} API. For example, consider a simple
query over the {\tt orders} table from the \mbox{TPC-H} benchmark:

\begin{small}
\begin{verbatim}
  SELECT orderkey, orderdate
  FROM orders 
  WHERE orderdate=`1996-01-01';
\end{verbatim}
\end{small}

\noindent This query would be expressed in Afterburner as follows:

\begin{small}
\begin{verbatim}
  sql.select()
    .field(`orderkey')
    .field(`orderdate')
    .from(`orders')
    .where(EQ(`orderdate', date(`1996-01-01')))
\end{verbatim}
\end{small}

\noindent Note that there is a very straightforward mapping from the
method calls to clauses in a standard SQL query, so we can view the
fluent API as little more than syntactic sugar. However, this query
API has a few advantages:\ First, it saves us from having to write a
query parser. Second, this fluent API is similar to
DataFrames~\cite{Armbrust_etal_SIGMOD2015}, an interface for data
manipulation that many data scientists are familiar with today.

Starting from an SQL query expressed in the fluent-style API,
Afterburner generates the string representation of the asm.js code
that corresponds to the query. In the current implementation, this is
performed in a straightforward way based on a small number of fixed
code templates in which various sub-expressions (e.g., the filter
predicate, join key, group by clause, etc.)\ are plugged. At present,
Afterburner has a fixed (hard-coded) physical plan for each class of
queries (i.e., it does not perform query optimization). Our prototype
implementation supports simple filter queries, inner joins, group bys,
simple aggregations, and order bys. We discuss below in more detail:

\smallskip \noindent {\bf Simple Filters.} A code template supports
generating query plans for simple filter--project or filter--aggregate
queries. The code template generates a loop that increments a record
iterator, which is used in combination with the starting offset of a
column to access a particular attribute. Inside the loop, the template
can either generate code to materialize a projection or to compute
simple aggregates such as {\tt COUNT}, {\tt AVG}, or {\tt SUM}.

\smallskip \noindent {\bf Joins.} The code template for supporting
filter--project or filter--aggregate queries over an inner join
implements a standard hash join. In the build phase, the code loops
over one relation to build the hash table. In the probe phase, the
generated code loops over the second relation to probe the hash table
for matching records, and then either materializes a projection or
computes an aggregate.

\smallskip \noindent {\bf Group Bys.} A group by plan loops over one
relation to build a hash table over one or more grouping keys. Another
loop is used to iterate over the hash table in order to process the
groups.

\section{Evaluation}
\label{sec:eval}

As indicated in the introduction, the main purpose of this work is to
demonstrate the {\it feasibility} of in-browser analytics. We freely
admit that Afterburner is far from a complete analytical database, but
we believe that our prototype is sufficient to answer this question.

In this section, we compare the performance of Afterburner with
vanilla JavaScript and MonetDB (v11.17.17) in exactly the same
execution environment on a commodity desktop machine. Afterburner and
vanilla JavaScript execute inside the browser, while MonetDB runs
natively.  For the vanilla JavaScript condition, we simply remove the
{\tt use asm} prologue directive so that generated code is not
validated as asm.js. However, the code still uses typed arrays.

For data, we used the TPC-H data generator with a scale factor of 1
GB:\ this creates a {\tt lineitem} table with 6~million rows and an
{\tt orders} table with 1.5~million rows. Table~\ref{exp:tab} shows
the four queries used in the evaluation. The first three queries are
very simple, and the final query is a slightly simplified version of
Q3 in the TPC-H benchmark.

\begin{figure*}[t]
  \includegraphics[width=0.98\textwidth, keepaspectratio]{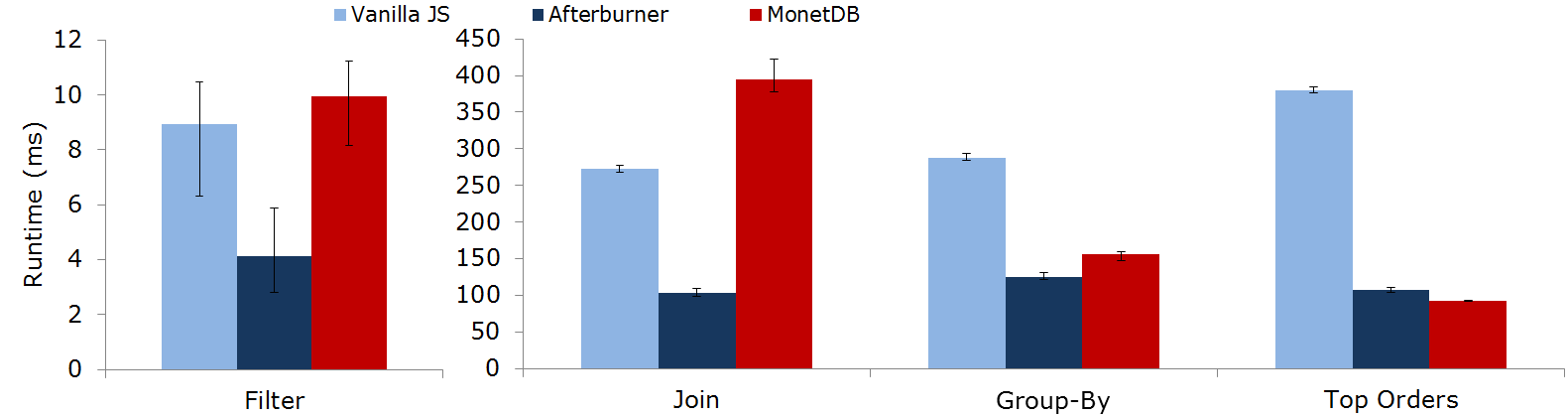}
  \caption{Query latency over five trials (with confidence intervals)
    for four sample queries, comparing vanilla JavaScript,
    Afterburner, and MonetDB. Both vanilla JavaScript and Afterburner
    run inside Firefox, while MonetDB runs natively on the same
    machine.}
  \label{fig:4q}
\end{figure*}

\begin{table}[t]
\begin{center}
\begin{tabular}{|p{0.9\columnwidth}|}
\hline
{\bf Q1: Filter} \\
\hline
\begin{minipage}{\columnwidth}
\vspace{0.2cm}
\begin{small}
\begin{verbatim}
SELECT count(*)
FROM orders
WHERE o_totalprice < 1500

\end{verbatim}
\end{small}
\end{minipage} \\
\hline
\hline
{\bf Q2: Join} \\
\hline
\begin{minipage}{\columnwidth}
\vspace{0.2cm}
\begin{small}
\begin{verbatim}
SELECT sum(o_totalprice)
FROM orders, lineitem
WHERE l_orderkey = o_orderkey

\end{verbatim}
\end{small}
\end{minipage} \\
\hline
\hline
{\bf Q3: Group-By} \\
\hline
\begin{minipage}{\columnwidth}
\vspace{0.2cm}
\begin{small}
\begin{verbatim}
SELECT o_orderdate, count(*)
FROM orders
GROUP BY o_orderdate

\end{verbatim}
\end{small}
\end{minipage} \\
\hline
\hline
{\bf Q4: Top Orders} \\
\hline
\begin{minipage}{\columnwidth}
\vspace{0.2cm}
\begin{small}
\begin{verbatim}
SELECT l_orderkey,
       sum(l_extendedprice) as rev,
       o_orderdate,
       o_shippriority
FROM orders, lineitem
WHERE l_orderkey=o_orderkey and
      o_orderdate between `1996-01-01'
                  and `1996-01-31'
GROUP BY l_orderkey, o_orderdate, o_shippriority
ORDER BY rev DESC LIMIT 10;

\end{verbatim}
\end{small}
\end{minipage} \\
\hline

\end{tabular}
\end{center}
  \vspace{-0.3cm}
  \caption{Sample queries used in our performance evaluation. Q4 is a
    slightly simplified version of Q3 from the TPC-H benchmark.}
  \label{exp:tab}
\end{table}

We ran performance evaluations on a commodity desktop with a six core
3.3~GHz AMD FX-6100 processor (6~MB of cache) and 16~GB of RAM,
running Ubuntu 12.04. Vanilla JavaScript and Afterburner ran in
Mozilla Firefox 43.0.4. We attempted to make the comparisons as fair
as possible: All cores were disabled except for a single one, since
code running inside a browser tab is single threaded. All our
performance measurements were on a warm cache---we first ran each query
five times, and then took measurements over the next five trials. For
Afterburner, the measured latency includes query compilation
overhead. In the case of vanilla JavaScript and Afterburner, all data
are explicitly loaded in memory; in the case of MonetDB, all data are
cached in the underlying OS buffer caches. This makes for a reasonably
fair comparison.

Evaluation results are shown in Figure~\ref{fig:4q}. We can see that
Afterburner is clearly faster than vanilla JavaScript due to the
asm.js optimizations. We find that the performance of Afterburner is
at least on par with, and in some cases exceeds the performance of
MonetDB. In the case of the join query Q2, examining the query plan
we find that MonetDB materializes the joined relation (all 6 million
rows) before counting them, and therefore is slower than Afterburner.

Admittedly, these results are somewhat surprising, but entirely
believable:\ Across a wide range of applications, a rough heuristic is
that asm.js runs in the browser at around half the speed of the same
application running natively. Since Afterburner uses compiled queries,
the reference point is a native C program. As Zukowski et
al.~\cite{Zukowski_etal_2005} have shown (e.g., Figure 3 in that
paper), there is still quite a bit of performance difference between
vectorized (but still interpreted) query execution (as with MonetDB)
and compiled queries.  So, it is plausible that AOT-compiled
JavaScript with proper optimizations (even with the overhead) still
runs faster than MonetDB---as our results suggest.  However, it is
worth emphasizing that in this evaluation, MonetDB is restricted to
running on a single core (for a fair comparison with single-threaded
JavaScript inside the browser), but this is an odd configuration for
analytical databases, which nearly always run on backend servers with
many-core processors.

In summary, these results show the feasibility of running an
analytical RDBMS completely inside the browser. Furthermore, the
performance of modern JavaScript engines is quite impressive---it's
ubiquity makes it an attractive platform for data analytics.

\section{So What?}
\label{subsec:scenb}

Having demonstrated the technical feasibility of an analytical RDBMS
in JavaScript, we turn our attention to the next obvious question:\ So
what? In this section, we discuss several answers.

A system like Afterburner can increase the speed and flexibility of
interactive data analytics. It is well known that the work of data
scientists, particularly in cases of exploratory tasks, is highly
iterative, where they ``poke'' at the data from many different
angles. Frequently, these tasks focus on a small subset of the
data---e.g., the data warehouse stores 180 days of log data, but the
data scientist is only interested in data from the last week. In this
case, the following describes some possible scenarios:
\begin{list}{\labelitemi}{\leftmargin=1em}
\setlength{\itemsep}{-2pt}

\item[1.] She repeatedly queries the entire dataset, but filters for
  the subset of interest. Achieving low query latency is entirely
  dependent on the analytical engine properly optimizing away unneeded
  work, which may be dependent on the physical storage of the data
  (how it's partitioned, compressed, etc.).

\item[2.] She materializes the data of interest:\ in an analytical
  database, this might involve selecting into a temporary table or
  creating a materialized view; in Spark, this might involve writing
  temporary data onto HDFS. Subsequent queries would be posed against
  this smaller dataset.

\item[3.] She materializes the data of interest and then copies it
  locally (e.g., to her laptop) for further manipulation. This might
  involve dumping data into a local RDBMS and then issuing additional
  queries or running Spark locally.

\end{list}

There are issues with all three approaches. In the first, the
analytics engine may not be smart enough to efficiently push down the
filters, especially for complex conditions. The second has a number of
drawbacks:\ In some cases, queries over the materialized data may no
longer run efficiently because the startup cost of the analytics
engine (e.g., Spark)\ dominates the actual processing time. In
analytical databases, creating materialized views might require
elevated privilege (not granted to everyone), and the ability to
create temporary tables presents additional data management
challenges. For example, who/what ``cleans up'' 50 tables named
variants of {\tt tmp}? Finally, the third approach creates an awkward
workflow and introduces friction. For an RDBMS, this involves copying
data over and ingesting locally before additional queries can be
run. Furthermore, this approach essentially requires maintaining two
separate analytics stacks. What if the version of custom
libraries running on the server becomes out of sync with the local
version?

Instead of the three above approaches, we propose interactive
analytics using something like Afterburner. If the backend analytics
engine is an RDBMS, the data scientist can transparently bring over
materialized data and continue querying without interrupting her
workflow, unlike in scenario (3) above. If the
backend analytics engine is something like Spark, we could easily
imagine building a bridge between Afterburner and
DataFrames. Furthermore, there's no principled reason why Afterburner
could not be extended to support Spark-style transformations and a
more imperative style of data analysis. In both cases, the data
scientist might be dealing with a relatively small amount of data, a
scenario that big data processing platforms do not optimize for (by
definition), and so Afterburner could end up with better performance
despite the overhead of JavaScript. Furthermore, if the data scientist
is already working in a browser-based notebook environment,
integration of Afterburner could be relatively straightforward.

\begin{table}[t]
\begin{center}
\begin{tabular}{|p{0.9\columnwidth}|}
\hline
{\bf Q5: Top Orders Variant} \\
\hline
\begin{minipage}{\columnwidth}
\vspace{0.2cm}
\begin{small}
\begin{verbatim}
SELECT l_orderkey, 
    sum(l_extendedprice * (1 - l_discount))
      as revenue, 
    o_orderdate, 
    o_shippriority 
FROM orders, 
    lineitem 
WHERE l_orderkey = o_orderkey 
   and o_orderdate = `1996-01-06'
GROUP BY l_orderkey, 
      o_orderdate, 
      o_shippriority
ORDER BY revenue
LIMIT 10

\end{verbatim}
\end{small}
\end{minipage} \\
\hline
\hline
{\bf Q6: Top Orders View} \\
\hline
\begin{minipage}{\columnwidth}
\vspace{0.2cm}
\begin{small}
\begin{verbatim}
SELECT l_orderkey, 
       sum(l_extendedprice * (1 - l_discount))
         as revenue, 
       o_orderdate, 
       o_shippriority 
FROM orders, 
       lineitem 
WHERE l_orderkey = o_orderkey and 
      o_orderdate between `1996-01-01'
                  and `1996-01-31' 
GROUP BY l_orderkey, 
         o_orderdate, 
         o_shippriority;

\end{verbatim}
\end{small}
\end{minipage} \\
\hline
  \end{tabular}
\end{center}
  \vspace{-0.3cm}
  \caption{Sample queries for the in-browser analytics scenario. We
    materialize the results of Q6 and filter on the client using
    Afterburner as an alternative to repeated queries of the type
    shown in Q5.}
  \label{exp:variants}
\end{table}

We sketch out what the performance of such an in-browser analytics
scenario might look like. Consider a TPC-H database at a 100 GB scale
factor:\ let's focus on the top orders query Q4 in
Table~\ref{exp:tab}, where the data scientist wants to dive into data
from January 1996. For the sake of argument, we assume this
exploration requires repeated querying that cannot be easily
captured in a single query.  We exemplify this with Q5 in
Table~\ref{exp:variants}, which pulls up the top 10 orders on a
particular day. Running this query in MonetDB takes 800~ms on a server
with dual 8 cores Intel Xeon E5-2670 processors (2.6~GHz) with 256~GB
of memory on Ubuntu 14.04. This corresponds to scenario (1)
above---run individual selective queries directly on the data
warehouse. Alternatively, the data scientist might materialize the
results of Q6 in Table~\ref{exp:variants} into a temporary table
(containing 2 million rows). Running a simple filter query for a
particular day on this temporary table (on the backend server) takes
28~ms. This corresponds to scenario (2) above, although we are not
taking into account network latency.

As an alternative, we can bring the materialized results of Q6 into
the browser and then run the filter query with Afterburner. Such a
query takes 25~ms on the same client machine described in
Section~\ref{sec:eval}. We do not include the data transfer time,
which would be amortized across multiple queries that probe the
materialized results. Such a scenario has the advantage of seamless,
low-latency client-side interactions (it's still SQL) without the data
management issues that come with creating temporary tables. Pushing
simple filter queries over to the client side naturally decreases load
on the server, and since Afterburner is just JavaScript, it can
integrate into any browser-based analytics tool, ranging from simple
dashboards to full-fledged interactive notebooks.

What's more, the ubiquity of JavaScript means that Afterburner can run
anywhere there is a browser, including mobile phones, tablets, and even the
connected toaster oven or refrigerator of the near
future. Java once promised developers that they can write their code
once and have it run anywhere, but JavaScript has come much closer to
actually delivering that promise---as evidenced by the range of
complex websites today that run on multiple platforms, on multiple
types of hardware, in multiple browsers.  This shows that analytics
capabilities in JavaScript can easily achieve widespread deployment.

Finally, we realize that our current working scenario---materializing
a subset of data for further analysis inside the browser---represents
a specific instance of split-execution strategies, where the client
(i.e., the browser) and the server collaborate to execute a query. In
our case, the user explicitly decides what queries run where, but this
need not be the case. For example, Franklin et
al.~\cite{Franklin_etal_SIGMOD1996} discuss {\it query shipping}
(basically, all systems today, where queries are shipped to the server
and results sent back), {\it data shipping} (where the server sends
the client tuples and query operators execute on the client), and {\it
  hybrid shipping} (where query operators execute on both the client
and the server). 
In our case, it could be possible for a query planner to automatically
place query operators either on the server or in the browser to
optimize some objective (latency, utilization, etc.). One of the
challenges with previous work is that the execution context on the
client and server might be different, but JavaScript nicely solves
this problem for us.

\section{Conclusion}

There is an emerging trend of building internet-scale services using
node.js, which unifies server-side and client-side processing around
JavaScript for customer-facing applications. Throw something like
Afterburner into the mix, perhaps we could integrate customer-facing
services with backend data analytics\ldots~all around JavaScript!
Such a thought should send shivers down the spine of any sane
developer, but this idea is so nutty we feel it's worth exploring!

\section{Acknowledgments}

We'd like to thank members of Waterloo's Data Systems Group (DSG) for
their encouragement (``you guys are nuts, but you should pursue this
anyway'') and Ashraf Aboulnaga for helpful comments on previous drafts
of this paper.


\begin{thebibliography}{1}

\bibitem{Armbrust_etal_SIGMOD2015}
M.~Armbrust, R.~S. Xin, C.~Lian, Y.~Huai, D.~Liu, J.~K. Bradley, X.~Meng,
  T.~Kaftan, M.~J. Franklin, A.~Ghodsi, and M.~Zaharia.
\newblock {Spark} {SQL}: Relational data processing in {Spark}.
\newblock In {\em Proceedings of the 2015 ACM SIGMOD International Conference
  on Management of Data (SIGMOD 2015)}, pages 1383--1394, Melbourne, Australia,
  2015.

\bibitem{Boncz_etal_CIDR2005}
P.~A. Boncz, M.~Zukowski, and N.~Nes.
\newblock {MonetDB/X100}: Hyper-pipelining query execution.
\newblock In {\em Proceedings of the Second Biennial Conference on Innovative
  Data Systems Research (CIDR 2005)}, pages 225--237, Asilomar, California,
  2005.

\bibitem{Franklin_etal_SIGMOD1996}
M.~J. Franklin, B.~T. {J\'{o}nsson}, and D.~Kossmann.
\newblock Performance tradeoffs for client-server query processing.
\newblock In {\em Proceedings of the 1996 ACM SIGMOD International Conference
  on Management of Data}, pages 149--160, Montreal, Quebec, Canada, 1996.

\bibitem{hiq}
K.~Krikellas, S.~Viglas, and M.~Cintra.
\newblock Generating code for holistic query evaluation.
\newblock In {\em Proceedings of the 26th International Conference on Data
  Engineering (ICDE 2010)}, pages 613--624, Long Beach, California.

\bibitem{Neumann11}
T.~Neumann.
\newblock Efficiently compiling efficient query plans for modern hardware.
\newblock {\em {PVLDB}}, 4(9):539--550, 2011.

\bibitem{Zaharia_etal_NSDI2012}
M.~Zaharia, M.~Chowdhury, T.~Das, A.~Dave, J.~Ma, M.~McCauley, M.~J. Franklin,
  S.~Shenker, and I.~Stoica.
\newblock {Resilient} {Distributed} {Datasets}: A fault-tolerant abstraction
  for in-memory cluster computing.
\newblock In {\em Proceedings of the 9th USENIX Symposium on Networked Systems
  Design and Implementation}, San Jose, California, 2012.

\bibitem{Zukowski_etal_2005}
M.~Zukowski, P.~Boncz, N.~Nes, and S.~{H\'{e}man}.
\newblock {MonetDB/X100}---a {DBMS} in the {CPU} cache.
\newblock {\em Bulletin of the IEEE Computer Society Technical Committee on
  Data Engineering}, 28(2):17--22, 2005.

\end{thebibliography}

\end{document}